\documentstyle[aps,prl,preprint]{revtex}
\begin{document}
\title
{Universal magnetic properties of frustrated quantum
antiferromagnets in two dimensions}
\author{Andrey V. Chubukov${}^{1,2}$, T. Senthil${}^{1}$ and
Subir Sachdev${}^{1}$}
\address{
${}^{1}$Departments of Physics and Applied Physics, P.O. Box 208284,\\
Yale University, New Haven, CT 06520-8284\\
and ${}^{2}$P.L. Kapitza Institute for
Physical Problems, Moscow, Russia}
\date{\today}
\maketitle
\begin{abstract}
We present a theory of frustrated, two-dimensional, quantum antiferromagnets
in the vicinity of a quantum transition from a non-collinear,
magnetically-ordered ground state to a quantum-disordered phase.
Using a sigma-model for bosonic, spin-1/2, spinon fields, we
obtain
universal scaling forms for a variety of observables.
Our
results are compared with
numerical data on the spin-1/2 triangular antiferromagnet.
\end{abstract}
\narrowtext

A useful classification of two-dimensional, quantum, Heisenberg
antiferromagnets
is provided by the structure of the magnetically-ordered ground state: the
spin-condensates on the sites can either be collinear or non-collinear
to each other. Collinear magnets have been extensively studied in recent
years and many of their properties are reasonably well understood. They possess
an $O(3)/O(2)$ order parameter whose fluctuations describe the
low temperature ($T$)
properties of the magnetically-ordered state~\cite{CHN}. The quantum-disordered
state has only integer spin excitations (the spinons are confined)
and spin-Peierls order is expected for
certain values of the single-site spin~\cite{Sach-Read1}.
The finite-$T$ crossover between these two states has also been
studied in some detail~\cite{CSY}.

Less is known, however, about non-collinear antiferromagnets, which
are the subject of this paper.
Examples include the triangular, kagome, and square (with first, second,
and third neighbor interactions) lattices. The magnetically-ordered state
completely breaks the spin-rotation symmetry, yielding an $SO(3)$ order
parameter~\cite{AM}. Space and time dependent twists of this order parameter
then define three independent spin-stiffnesses, spin-susceptibilities,
and associated spin-wave velocities. For simplicity, we will restrict
our attention here to magnets with {\em coplanar\/} spins and an internal
symmetry (a $C_{3v}$ symmetry on the triangular and kagome lattices, and a
screw axis symmetry for the incommensurate planar spirals on the square
lattice), which leads to just two independent stiffnesses ($\rho_{\perp}$,
$\rho_{\parallel}$), susceptibilities ($\chi_{\perp}$, $\chi_{\parallel}$),
and spin-wave velocities ($c_{\perp} = (\rho_{\perp}/\chi_{\perp})^{1/2}$,
$c_{\parallel} = (\rho_{\parallel}/\chi_{\parallel})^{1/2}$); more complicated
non-collinear magnets will have similar properties.
The long-wavelength action for the $SO(3)$ order parameter
has an $SO(3) \times O(2)$ symmetry, the $O(2)$ being a
continuum manifestation
of the internal symmetry noted above~\cite{AM}.
A spacetime dimension $D=2+\epsilon$
study of small fluctuations of the $SO(3)$ order parameter about the
magnetically-ordered state was performed by Azaria {\em et.
al.\/}~\cite{Azaria};
they found that the stiffnesses and susceptibilities became asymptotically
equal upon approaching the critical point separating the
magnetically-ordered and quantum-disordered phases, with the critical theory
possessing an enlarged $O(4)$ symmetry. A large $N$ theory based upon
$Sp(N)$ symmetry~\cite{Sach-Read2} found a similar magnetically-ordered state,
but was also able to access the quantum-disordered phase. The latter
state was predicted to be a featureless, fully gapped spin-fluid, with
unconfined, bosonic spin-1/2 spinon excitations.
We also note that there are alternative approaches
to the quantum disordered phase~\cite{Kalm_Laugh}
which are quite disconnected
from the structure of the ordered state.

In this paper, we shall present a theory of the universal,
finite-$T$ properties
of non-collinear antiferromagnets in the vicinity of the critical point.
We will describe the crossover from the magnetically-ordered state,
with its low-lying spin-wave excitations,
to the fully gapped quantum-disordered state
via an intermediate quantum-critical region. Our results are in complete
agreement with some previous studies of the magnetically-ordered
state~\cite{Azaria}
and the quantum disordered state~\cite{Sach-Read2},
and establish a fundamental connection
between the $O(4)$-symmetric critical point of Ref.~\cite{Azaria} and
the deconfined bosonic spinons of Ref.~\cite{Sach-Read2}; a related
connection was noted recently in Ref.~\cite{Aza}.
We will also obtain new results for the low $T$ behavior of the dynamic
structure factor and uniform susceptibility of magnetically-ordered
antiferromagnets.

Our motivation for this study is similar to that for the anolagous recent
study of collinear antiferromagnets~\cite{CSY}.
A given $S=1/2$
antiferromagnet may be either magnetically-ordered (as is expected
for the triangular lattice) or
quantum-disordered (the kagome lattice)~\cite{And-Fasek}.
At low $T$, the magnetically ordered magnet has thermally-excited
classical spin-wave fluctuations (the renormalized-classical (RC) region),
while the quantum-disordered magnet has only activated deviations from its
ground-state properties. At higher $T$ however both these magnets are expected
to crossover to a quantum-critical~\cite{CHN} (QC) region
where classical and thermal
fluctuations are equally important. Many properties of this region
are universal, and are thus amenable to numerical and experimental tests.
In particular, there are significant quantitative differences
between the QC behavior of collinear and non-collinear
magnets, which are a direct consequence of the presence of deconfined spinons
in the latter.

We begin by presenting our effective action. We choose to describe
the local spin configuration by an $SU(2)$ rotation about a reference ordered
state. The choice of $SU(2)$ rather than $SO(3)$ is significant, and has the
immediate consequence of suppressing the vortices~\cite{Kaw-Miyash}
associated with $\pi_1 ( SO(3))=
Z_2$ for which the $SU(2)$ field is double-valued. This choice is motivated
partly by the results of Ref.~\cite{Sach-Read2}, where vortices were suppressed
in the quantum-disordered phase by a Higgs condensate. We parametrize the
$SU(2)$ matrix by two complex numbers $z_1$, $z_2$ with $|z_1|^2 + |z_2|^2 =1$,
and write down the most general, long-wavelength action with an $SU(2) \times
O(2)$ invariance:
\begin{equation}
{\cal S} =
\int d^2 x d \tau \frac{1}{g_{\mu}}
\left[ \partial_{\mu} z^{\dagger} \partial_{\mu} z
- \frac{\gamma_{\mu}}{4}
\left( z^{\dagger} \partial_{\mu} z - \partial_{\mu}
z^{\dagger} z \right)^2 \right]
\label{C1}
\end{equation}
It is easy to show that
$g_x = 1/2 \rho^{0}_{\perp}, g_{\tau} = 1/2\chi^{0}_{\perp},
\gamma_x = (\rho^{0}_{\parallel} - \rho^{0}_{\perp})/\rho^{0}_{\perp},
\gamma_{\tau} = (\chi^{0}_{\parallel} - \chi^{0}_{\perp})/\chi^{0}_{\perp}$,
where the superscript 0 denotes bare values; note that if the $\gamma_{\mu}=0$,
${\cal S}$ has an enlarged $O(4)$ symmetry.
The action ${\cal S}$ can
be explicitly derived by a long-wavelength analysis of
the models of Refs.~\cite{Azaria}
and~\cite{Sach-Read2}; we have also learned of a recent study of ${\cal S}$
by Azaria {\em et. al.\/}\cite{recent}. The staggered spin-structure factor
(wavevectors measured as deviations from the ordering wavevector $\vec{G}$)
can be shown to be the Fourier transform of
$\mbox{Re} \left( z^{\dagger} (x_1 , \tau_1 ) z (x_2 , \tau_2 ) \right)^2$.
Note that this is {\em quartic} in the $z$, consistent with the identification
of the $z$ quanta as spin-1/2 bosonic spinons.

We studied ${\cal S}$ by generalizing $z$ to an
$N$-component, unit-length, complex vector, and performing a $1/N$
expansion; ${\cal S}$ then has a
$SU(N)\times O(2)$ invariance, while for $\gamma_{\mu}=0$ it is invariant
under $O(2N)$.
This method allows us to work directly in $D=2+1$ and access both
the QC
and RC regions.
Note that the extension to large $N$
is different from that used in Refs.~\cite{Aza,Pfeuty}.

We expect that ${\cal S}$ possesses quantum-disordered and magnetically-ordered
(with the $z$ quanta condensed)
as the couplings (say $g_x$) are varied.
A key property of the present large $N$ expansion is that the long-distance
physics at the
critical point at $g_x = g_c$ is $O(2N)$-symmetric.
 This is manifested
in the magnetically ordered phase ($g_x < g_c$) by the critical behavior of the
stiffnesses. Josephson scaling is obeyed by the fully renormalized
$\rho_{\parallel}$,
$\rho_{\perp}$, $\chi_{\parallel}$, $\chi_{\perp}$
all of which vanish as
$(g_c - g_x)^{\nu}$, where $\nu$ is the correlation length exponent
($\nu = 1 - 16/3 N \pi^2  + {\cal O} (1/N^2 )$). However, the relative
differences between the stiffnesses also vanish at the critical point:
we defined $\Delta_1 = (\rho_{\parallel} - \rho_{\perp})/\rho_{\perp}$,
$\Delta_2 = (\chi_{\parallel} - \chi_{\perp} ) / \chi_{\perp}$, and found
\begin{eqnarray}
\Delta_1  &=&  \gamma_{1} (\xi_{J})^{-\phi_1} + \gamma_2
(\xi_J)^{-\phi_2} \nonumber\\
\Delta_2  &=&  \gamma_{1} (\xi_{J})^{-\phi_1} - 2\gamma_2
(\xi_J)^{-\phi_2}
\label{C14}
\end{eqnarray}
where $\gamma_1 = (2 \gamma_x + \gamma_{\tau})/3; ~\gamma_2 = (\gamma_x -
\gamma_{\tau})/3$, and $\xi_J$ is
the Josephson length measured in lattice units. The
positive crossover exponents
$\phi_{1,2}$ measure the irrelevancy of the
$\gamma_{\mu}$ terms in ${\cal S}$;
the $\gamma_{\mu}$ are actually `dangerously'-irrelevant as $\Delta_{1,2}$
control long-wavelength physics for $g_x < g_c$.
To order $1/N$, we found $\phi_1 =  1 + 32/3 \pi^2 N$, $\phi_2 = 1+112/15 \pi^2
N$~\cite{Lang_Ruhl}.

We now present our scaling
results for the wavevector- ($k$) and frequency- ($\omega$) dependent
staggered ($\chi_s$) and uniform ($\chi_u$)
spin susceptibilities in the vicinity of $g_x = g_c$.
We restrict ourselves to
$g_x < g_c$, although more complete results have been
obtained~\cite{long}. We found
\begin{eqnarray}
\chi_s (k ,\omega) &=&
\frac{2\pi N^{2}_0}{N \rho_{\perp}} \left(\frac{\hbar c_{\perp}}{k_B T}
\right)^2
 \left(\frac{N k_B T}{4 \pi \rho_{\perp}}\right)^{\bar \eta}
\Phi_s \left(\overline{k}, \overline{\omega}, x, \Delta_1, \Delta_2 \right)
\nonumber \\
\chi_u (k ,\omega) &=& \left(\frac{g \mu_B}{\hbar c^{2}_{\perp}}\right)^2
{}~k_B T ~\Phi_u \left(\overline{k},
\overline{\omega}, x,  \Delta_1, \Delta_2 \right)
\label{I30}
\end{eqnarray}
where $N_0$ is the on-site magnetization at $T=0$,
$\Phi_{1s}$,  $\Phi_{1u}$
are universal functions of the dimensional
variables $\overline{k} = \hbar c_{\perp} k/ k_B T$, $\overline{\omega} =
\hbar \omega /k_B T$, $x = N k_B T/ 4 \pi \rho_{\perp}$.
We found the exponent $\bar{\eta} =
1 + 32/3 \pi^2 N$. The prefactor of $\Phi_s$ remains non-singular at
$g_x = g_c$ as $N_0 \sim (g_c - g_x)^{\bar{\beta}}$ with $2\bar{\beta} =
(1 + \bar{\eta})\nu$.
All scaling functions are defined such that they
remain finite as $x \rightarrow \infty$.
As before~\cite{CSY},
the argument $x$ determines whether the system is in
the QC ($x \gg 1$) or RC
($x \ll 1$) region.

An important difference in the above scaling forms from those for collinear
magnets~\cite{CSY}
is in the value of $\bar{\eta}$. Here we have $\bar{\eta}$ close to unity,
while the analogous exponent for collinear magnets was close to zero.
This is a consequence of the presence here of deconfined spinons:
it is the $z$ quanta which behave like almost free particles
(at $T=0$,
$\langle z^{\dagger} z \rangle \sim 1/k^{2-\eta}$ with $\eta$ close to $0$)
while the staggered susceptibility is a correlator of a composite operator
of two spinons
($\chi_s \sim 1/k^{2-\bar{\eta}}$ with $\bar{\eta}$ close to 1).

We have computed $\Phi_s$, $\Phi_u$ in a $1/N$ expansion to linear order
in $\Delta_{1,2}$. We
describe our results as they relate to various observables.

{\em Correlation length.}
As in collinear magnets, we define the correlation length, $\xi$,
from the long-distance $e^{-r/\xi}$ decay of the equal-time spin-spin
correlation function. We found that, to order $1/N$, there is a
simple relationship between the values of $\xi$ for collinear and
non-collinear magnets. For all values of $x$,
the non-collinear $\xi$ is precisely $1/2$ the previously computed
$\xi$~\cite{CSY}
for the {\em isotropic} $O(2N)$ sigma model. The factor of $1/2$
is a signature of deconfined spinons. The collinear expression
for $\xi$~\cite{CSY}
however must be used with the effective values $\rho_s =
\rho_{\perp} ( 1 + N \Delta_1 / (2 N^2 - 2) )$, $\chi = \chi_{\perp}
( 1 + N \Delta_2 / (2 N^2 -2 )$ and $c= (\rho_s / \chi)^{1/2}$; notice also
the factor of $4$ difference in the couling constant in (\ref{C1}) and in
{}~\cite{CSY}.
For the physical case $N=2$, we have to first order in $\Delta_{1,2}$
that $\rho_s = (2 \rho_{\perp} + \rho_{\parallel})/3$, $c=
(2 c_{\perp} + c_{\parallel})/3$, and our result for $\xi$ is then
consistent in the RC region with that of Azaria {\em
et.al.\/}~\cite{Azaria}.

{\em Static uniform susceptibility}.
The result for $\chi_u$ is obtained by evaluating the response to a
vector potential coupled to the conserved charge of the $SU(N)$ symmetry.

In the RC region ($N k_B T \ll 4 \pi \rho_s$)
we obtained
\begin{equation}
\chi_{u} = \left(\frac{g \mu_{B}}{\hbar}\right)^{2}
{}~\left(\frac{(N \chi_{\perp} +
\chi_{\parallel})}{(N+1)\chi_{\perp}}\right)~\left[\frac{2 \chi_{\perp}}{N} +
\frac{N-1}{N}~
\frac{k_{B} T}{2 \pi c^{2}}\right]
\label{C22'}
\end{equation}
It is worth emphasizing
 that although we are considering an essentially classical regime, the
$T$ dependence of $\chi_u$ is a purely quantum effect - it
disappears if the spin-waves had a classical, thermal distribution.

In the QC region ($N k_B T \gg 4 \pi \rho_s$),
we found to order $1/N$
\begin{equation}
\chi_{u} = \left (\frac{g \mu_{B}}{\hbar c}\right)^{2}
{}~k_B T \frac{\sqrt{5} \Theta}{4\pi}~\left[\left(1 - \frac{0.31}{N}\right) +
\alpha~\bar{x}^{-1/\nu} +\ldots\right]
\label{Q1}
\end{equation}
where $\Theta = 2 \log[(\sqrt{5}+1)/2]$,~ $\bar{x} = N k_B T/4 \pi \rho_s$
and $\alpha = 0.8 + O(1/N)$.
Note that the slope of the linear in $T$ term is precisely
1/2 of that in the $O(2N)$ sigma-model~\cite{CSY}.
 The factor of $1/2$ is again a signature of
spin-1/2 spinons and should be amenable to experimental tests.

{\em Staggered dynamic susceptibility and structure factor}.
In the RC region,
the scaling form (\ref{I30}) for $\chi_s$ collapses into a reduced
scaling form in which the physical $\xi$, rather than $c/k_B T$ is
the most important length scale~\cite{CHN,CSY}. We found
\begin{eqnarray}
\chi_s (k, i\omega_n) &=& \frac{N^{2}_0}{\rho_s (N-1)}
 \left[\frac{k_B T (N-1)}{4 \pi \rho_s}\right]^{(N+1)/(N-1)} \times \nonumber
\\
&& ~\xi^2 ~ f(k \xi, \omega_n \xi/c)
\label{C38}
\end{eqnarray}
were $f$ is a scaling function.
Note that computations were in fact done only to order $1/N$ - the
form at arbitrary $N$ follows from a reasonable guess about
the wavefunction renormalization of the composite field.
The overall factor in (\ref{C38}) is chosen such that $f(0,0) =1 +
{\cal O}(1/N)$,
The behavior of $f(x,y)$ at intermediate $x,y ={\cal O}(1)$ is rather
complicated, chiefly
because spin-wave velocity also acquires a substantial downturn
renormalization at $k\xi ={\cal O}(1)$~~\cite{CHN}. However
 at $k \xi \sim \omega \xi/c \gg 1$, velocity renormalization is
irrelevant and we obtained
\begin{equation}
f(x,y) = \left(\frac{N-1}{N +1}\right)~\frac{1}{x^2 + y^2}~
\left[\frac{\log(x^2 + y^2)}{2}\right]^{(N+1)/(N-1)}
\nonumber
\end{equation}
It is not difficult to demonstrate that this result for $f (x,y)$
yields a $\chi_s (k,
\omega)$ which is precisely the
rotationally-averaged spin-wave result  for
the ordered antiferromagnet, as it of course should be at $k \xi \gg
1$ but $k \xi_J \ll 1$.

We also computed $\mbox{Im} \chi_s ( k, \omega)$ for real $\omega$.
In the RC
region, we describe the results using the dynamic structure factor
$S(k, \omega)$ which satisfies
\begin{equation}
S(k, \omega) = N^{2}_{0}~ \left(\frac{ k_B T}{\rho_s}\right)^2
{}~\left( \frac{\bar{\Xi} (k, \omega)}{\omega}\right)
\label{C43}
\end{equation}
where $\bar{\Xi}$ is straightforwardly related to
$\Phi_s$ introduced in (\ref{I30}). For experimental comparisons, it
is sufficient to consider the frequency range $\omega \leq c/\xi$. We then
found  $\bar{\Xi} (k, \omega) = \omega/ 2\pi c k^3$ for $c k \gg \omega$,
(in this region of $k$, collisionless Landau damping is dominant),
and  $\bar{\Xi} (k, \omega) \propto  (\omega \xi^3 /c)~\left((N-1) k_B
T /4 \pi \rho_s \right)^{(5-N)/2 (N-1)}~$  for $c k \sim \omega$
(the dominant contribution is the damping of quasiparticles).

Now the QC region. Here we  restrict our results  to the
critical point $x=\infty$, and negligible anisotropy ($\Delta_{1,2} = 0$).
For $\hbar c k, \hbar \omega \gg k_B T$ we obtained
\begin{equation}
\Phi_s = \frac{A_{N}}{16
(\overline{k}^{2} - (\overline{\omega} + i \delta)^2)^{1 - \bar{\eta}/2}}
\label{new}
\end{equation}
where
$A_{N} = 1 +{\cal O}(1/N)$.
At small $k$ and $\omega$, we have $\mbox{Re}\Phi_s =$
$(\sqrt{5}/16 \pi \Theta)[1 -$ $(\overline{k}^{2}(1 + 2 \Theta/\sqrt{5}) -$ $
\overline{\omega}^{2})/12 \Theta^2\/$ $+ \ldots]$ where
 $\ldots$ stand for higher powers
of $\overline{k}, \overline{\omega}$ and for regular corrections in $1/N$.
For $\mbox{Im}\Phi_s$
we obtained
the following asymptotic limits for large $N$
\begin{displaymath}
\mbox{Im} \Phi_s =\left\{
\begin{array}{l} \displaystyle
\frac{A_N \sin(\pi \bar{\eta}/2)}{16}~\frac{\theta (\overline{\omega}^2 -
\overline{k}^2)}{(\overline{\omega}^2 - \overline{k}^2)^{1 -{\bar \eta}/2}}
{}~;~ \overline{\omega}\gg 1 \\ \displaystyle
 \frac{\overline{\omega}}{8
\sqrt{\pi}}~\frac{e^{-\overline{k}/2}}{\overline{k}^{3/2}}~;~\overline{\omega}
\ll 1,~~\overline{k}\gg 1
\end{array}\right.
\end{displaymath}
where $\theta
(x)$ is a step function. In both cases, Landau damping is dominant.
Finally, when  both $\overline{\omega} \ll 1$ and $\overline{k} \ll 1$,
 quasiparticle excitations are overdamped and we only know that
$\mbox{Im} \Phi_s \propto \overline{\omega}$.

In the $T=0$ quantum-disordered phase, $\mbox{Im} \Phi_s$
shows a clear signature of deconfined spinons - the spectral weight
at fixed $k$ is a broadband continuum rather than the delta-function
peak present in collinear magnets~\cite{long}.

{\em Local susceptibility and spin-lattice relaxation}.
The local dynamic structure factor  $S (\omega)$ is given by
 $ S(\omega) = \int d^2 k ~S (k, \omega)/ 4 \pi^2$.
Simple inspection then shows that for $\omega \sim
c \xi^{-1}$, the integration over momentum is also
confined to $k \sim \xi^{-1}$ and therefore $S(\omega)$ is a universal
observable. The small frequency limit of $S(\omega)$ is directly related to the
spin-lattice relaxation rate of
nuclear spins coupled to electronic spins in the antiferromagnet:
$1/T_1 \propto~S(\omega \rightarrow 0)$. In the RC region,
 we find using our previous results for the scaling functions
that for $\omega \sim c \xi^{-1}$
\begin{equation}
S (\omega) \propto \frac{N^{2}_{0} \xi}{c}~ \left(\frac{(N-1) k_B
T}{4 \pi \rho_s}\right)^{(3N +1)/2(N-1)}
\label{C49}
\end{equation}
For $N=2$, we then have  $1/T_1 \propto T^{7/2}~\xi$.

Deep in the QC region, we found
\begin{equation}
S(\overline{\omega}) = \frac{4 \pi \hbar N^{2}_0}{N \rho_{\perp}}
 ~~\left(\frac{N k_B T}{4
\pi \rho_{\perp}}\right)^{\bar \eta}
{}~\frac{K (\overline{\omega})}{1 - e^{-\overline{\omega}}}
\label{I35}
\end{equation}
where   $K (\overline{\omega}) = \overline{\omega}~B_N
\sin(\pi \bar{\eta}/2)~/32 \pi$
at  $\overline{\omega} \gg 1$, and
$K (\overline{\omega}) = \overline{\omega}~C_N (\sqrt{5} -1)/ 64\pi$
at  $\overline{\omega} \ll 1$. The factors $B_N$ and $C_N$ both behave as
$1 + {\cal O}(1/N)$. Clearly then, $1/T_1 \propto T^{\bar \eta}$.

{\em Static structure factor.}
Unlike collinear magnets,
the static structure factor
$S(k) = \int d \omega  ~S (k, \omega)~/2\pi$ in non-collinear
antiferromagnets
is nonuniversal because the frequency integral over quantum
fluctuations is divergent. This follows from the behavior
at large frequencies where $S (k, \omega) \propto
1/ \omega^{2-{\bar \eta}}$ and ${\bar \eta} >1$.
The nonuniversality is however more relevant for the
QC region, where $T$ is the only scale
for fluctuations; in the RC region,
$\xi$ is
exponentially large, and there is a universal contribution to
$S(k)$ from classical fluctuations which scales as $\xi^2$.
 In the RC region we then have
 $S(k) \approx k_B T \chi_s (k,0)$,
where $\chi_s (k, 0)$ is given by (\ref{C38}). At $k=0$ this yields
  $S(0) \propto T^{2 N/(N-1)}~\xi^2$.
For $N=2$, we then obtain $S(0) \propto T^4 \xi^2$.

{\em Application to the $S=1/2$ triangular antiferromagnet}.
We performed a $1/S$ expansion on this antiferromagnet
to obtain the $T=0$ values of $\rho_{\perp}$, $\rho_{\parallel}$,
$\chi_{\perp}$, $\chi_{\parallel}$ (all to order $1/S$),
and $N_0$ (to order $1/S^2$).
For $S=1/2$ this gave us $N_0 = 0.266$; $\chi_{\perp} = 0.07/J a^2$;
$\chi = 0.077/J a^2$; $\rho_s = 0.087J$; $c = (\rho_s /\chi)^{1/2} = 1.06J a$.
For the uniform susceptibility in the RC regime
 we then obtained
$\chi_{u} = \left(g \mu_{B}/\hbar a \right)^2$ $[
0.08/J +$ $0.08 k_{B} T/J^2 +$ ${\cal O}(T^2/J^3)]$.
On the other hand, in the QC regime, we have
$\chi_{u} = \left(g \mu_{B}/\hbar a \right)^2$ $[
0.13 k_{B} T/J^2 +$ $0.07/J~ (k_B T/2\pi \rho_s)^{1-1/\nu} + \ldots]$.
The temperature dependence in the subleading term is likely to be quite small
in the region of experimental interest ($k_B T \sim 2 \pi \rho_s$), and we
can well approximate this term by a constant. Note however, that the factor
0.07 is an $N=\infty$ result - the $1/N$ corrections to this factor
have not been computed.
Further, the correlation length behaves in the RC regime
as $\xi \approx 0.25 \left (4 \pi \rho_s /k_B T \right)^{1/2}~
\exp[4 \pi \rho_s /k_B T]$
where $4 \pi \rho_s \approx 1.09 J$, and
deep in the QC region as $\xi = 0.53 J a /k_B T$.

Consider now the numerical results for $\chi_u$.
 The data of recent series expansion studies~\cite{Young} show
that $\chi_u$ obeys a Curie-Weiss law at high
$T$, passes through a maximum at $T \approx 0.4J$, and then falls down.
The region below the maximum is quite small; nevertheless,
we fitted this data by a straight line
and found $0.13 \pm 0.03$ for the slope and about $0.06$
 for the intercept - the results are in better agreement with
our QC rather than the RC result.
Finally, at very low $T$, we expect a crossover to the RC
regime, and the corresponding value of $\chi_u$ at $T=0$
is also consistent with the data.
We also compared the data for
the correlation length and $S(0)$ at $k_B T \sim 0.4J$
and found rough consistency with our expressions
in the crossover region between QC and RC
regimes. Note that our interpretation of the numerical data is
different from that in Ref.~\cite{Young}.

To conclude,
we have presented a theory of the critical properties of non-collinear
quantum antiferromagnets in two dimensions. Our key assumption was
on the validity of a continuum description in $SU(2)$ variables, which
suppressed vortex excitations. However, we were then able to show that
our results were consistent with earlier large $N$~\cite{Sach-Read2} and
$D=2+\epsilon$~\cite{Azaria} studies.
The quantum disordering transition was described
by an
anisotropic sigma-model for spin-1/2, bosonic spinon fields.
All physical observables involve a collective mode of two spinons, and
we computed explicit scaling forms for a variety of experimentally
measurable quantities.
Our results for $\chi_u$ in the QC region are
roughly consistent with
recent numerical data on the spin-1/2 triangular antiferromagnet~\cite{Young};
this may be viewed as indirect evidence for the presence of deconfined
spinons.
However, numerical results
also seem to indicate that the $T$ range where QC behavior
may be observed is rather narrow for this system.
More detailed studies, especially
in quantum-disordered non-collinear magnets will be quite useful.

The research was supported by NSF Grant No. DMR-9224290.
S.S. is grateful for LPTHE, Universit\'{e} Paris 7, for hospitality.
We thank
P. Azaria, B. Delamotte, P. Lechmenniat, D. Mouhanna and N. Read
for helpful discussions.

\end{document}